\documentclass[12pt]{article}

\usepackage{amsmath,amssymb,cite,bm}
\usepackage{graphicx}

\makeatletter
 
  \@addtoreset{equation}{section}
 \makeatother

\newcommand{\be}{\begin{equation}}
\newcommand{\ee}{\end{equation}}
\newcommand{\bea}{\begin{eqnarray}}
\newcommand{\eea}{\end{eqnarray}}
\newcommand{\beann}{\begin{eqnarray*}}
\newcommand{\eeann}{\end{eqnarray*}}

\newcommand{\ba}{\begin{array}}
\newcommand{\ea}{\end{array}}

\allowdisplaybreaks

\begin{document}

\setlength{\oddsidemargin}{0cm}
\setlength{\baselineskip}{7mm}

\begin{titlepage}
\renewcommand{\thefootnote}{\fnsymbol{footnote}}
\begin{normalsize}
\begin{flushright}
\begin{tabular}{l}
UTHEP-669
\end{tabular}
\end{flushright}
  \end{normalsize}

~~\\

\vspace*{0cm}
    \begin{Large}
       \begin{center}
         {Matrix Geometry and Coherent States}
       \end{center}
    \end{Large}
\vspace{0.7cm}

\begin{center}
Goro I{\sc shiki}$^{ }$\footnote
            {
e-mail address : 
ishiki@het.ph.tsukuba.ac.jp}
      
\vspace{0.7cm}
                    
     $^{ }$ {\it Graduate School of Pure and Applied Sciences, University of Tsukuba, }\\
               {\it Tsukuba, Ibaraki 305-8571, Japan}\\

     $^{ }$ {\it Yukawa Institute for Theoretical Physics, Kyoto University}\\
               {\it Kyoto, 606-8502, Japan}\\

\end{center}

\vspace{0.7cm}

\begin{abstract}
\noindent
We propose a novel method of finding 
the classical limit of the matrix geometry.
We define coherent states 
for a general matrix geometry described by 
a large-$N$ sequence of 
$D$ Hermitian matrices $X_{\mu} \; (\mu =1,2, \cdots, D)$ 
and construct a corresponding classical space 
as a set of all coherent states.
When the classical space forms a smooth manifold,
we also express various geometric 
objects on the classical space 
such as the metric, Levi-Civita connection, curvature and 
Poisson tensor, in terms of the matrix elements.
This method provides a new class of observables 
in matrix models, which characterize 
geometric properties of matrix configurations.

\end{abstract}
\vfill

\end{titlepage}
\vfil\eject

\setcounter{footnote}{0}


\section{Introduction}
Matrix models are conjectured to give nonperturbative formulations 
of string and M theories\cite{Banks:1996vh, Ishibashi:1996xs}. 
In the matrix models, classical Riemannian geometry 
is replaced by a kind of quantum geometry described by 
matrices\cite{deWit:1988ig} 
and the matrix models are expected to realize
a novel description of gravitational theories based on the matrix geometry.
However, the relation between the matrix geometry
and Riemannian geometry has not been fully understood so far
\footnote{
For infinitely large matrices, a very interesting solution to this problem
was found in \cite{Hanada:2005vr}. }.
Finding a clear relation between these two geometries, which 
look quite different at first sight, 
 will help to understand the matrix models as theories of gravity.

Although concrete examples of the matrix geometry (the matrix regularization) 
have been constructed for some 
manifolds\cite{Madore:1991bw,Connes:1997cr,Balachandran:2001dd,SheikhJabbari:2005mf,Arnlind:2006ux}, any algorithmic 
construction method has not been known for a general manifold. 
Also, its inverse problem, namely the problem of 
finding an associated classical (commutative) manifold for a 
given matrix geometry, has been only partly understood.
For the latter problem, the Morse theoretic method\cite{Shimada:2003ks} 
and the method of taking an optimal gauge\cite{Hotta:1998en,Azeyanagi:2009zf}
have been proposed. These are useful to see 
some geometric aspects such as the topology or the shape of 
the classical space for a given matrix configuration.
They are, however, not convenient enough to see more 
detailed information such as
relations between matrix configurations and 
geometric objects on the classical space such as the curvature tensors.
For the case of two-dimensional surfaces, 
matrices which correspond to some geometric 
objects were explicitly constructed in \cite{Arnlind:2010ac}.
Recently, another method of finding the classical space 
using a Dirac operator was proposed in \cite{Berenstein:2012ts,Berenstein:2013tya}. 
This method is closely related to our proposal.

In this paper, we propose a novel approach to the latter
problem by generalizing the notion of coherent states. 
Coherent states are often considered for the Heisenberg algebra 
$[\hat{p}, \hat{q}]=-i\hbar $ (see Appendix \ref{Canonical coherent states})
and play an important role in taking the classical limit.
They are approximate simultaneous eigenstates of
$\hat{p}$ and $\hat{q}$ and 
can be defined as the states which saturate the uncertainty bound.
They are considered as the quantum analogue of the 
points on the classical phase space.
Indeed, there exists a one-to-one correspondence between 
the classical plane and
a set of all coherent states.
This means that, from the Heisenberg algebra,
one can construct the classical geometry as a set of all coherent states.
The same construction is possible for the case of the fuzzy sphere, where 
the Bloch (spin) coherent states are in one-to-one 
correspondence with the points on $S^2$ (see Appendix 
\ref{Bloch coherent states}).
In this paper, we extend this construction to more general matrix geometries.

For a Poisson manifold ${\cal M}$, its matrix regularization 
can be defined in terms of a large-$N$ sequence of matrices 
(for example, see \cite{Arnlind:2010ac}).
Let $T_N$ be a linear map from $C^{\infty}({\cal M})$ to 
a set of all $N \times N$ matrices.
When the sequence $\{T_N\}_N $, where $N$ is monotonically 
increasing to infinity, satisfies the four conditions shown in 
Eqs. (4.1)-(4.4) in \cite{Arnlind:2010ac}, $\{T_N\}_N $ 
is called the matrix regularization of ${\cal M}$.
The main property of the matrix regularization is the first two of the four,
which take the following form:
\begin{align}
&\lim_{N \rightarrow \infty} || T_N(f)T_N(g)-T_N(fg) || =0, 
\nonumber\\
&\lim_{N \rightarrow \infty} || iN[T_N(f), T_N(g)]-T_N(\{f,g \}) || =0.
\label{def of MR}
\end{align}
These conditions say 
that the map preserves the algebraic structure of functions and 
the Poisson bracket is well approximated by the commutators of matrices.
In this paper, we consider the case where the manifold ${\cal M}$ is 
embedded in a flat space. 
For the embedding function $y^{\mu}:{\cal M}\rightarrow {\bf R}^D$,
we denote by $X^{\mu}=T_N(y^{\mu})$ the images of $y^{\mu}$.
Then, the first condition in (\ref{def of MR}) means that the polynomials 
$y^{\mu}y^{\nu}\cdots $ are mapped to the matrix polynomials 
$X^{\mu}X^{\nu}\cdots $ up to the $1/N$ corrections, while 
the second means that 
\begin{align}
iN [X^{\mu}, X^{\nu}] - W^{\mu\nu}(X)\rightarrow 0,
\label{commutator of XX}
\end{align}
where $W^{\mu \nu}$ is the induced Poisson tensor\footnote{
Of course, there is an ordering ambiguity in defining 
$W^{\mu \nu}$ with the matrix argument. 
However, (\ref{commutator of XX}) also 
implies that $X^{\mu}$ commute with each other in 
the large-$N$ limit so that the ambiguity is negligibly small.}.
Thus, the matrix regularization for an embedded space is given by 
a large-$N$ sequence of Hermitian matrices $\{X^{\mu}\}_N$
which satisfies (\ref{commutator of XX}).
See \cite{Arnlind:2006ux} for more detailed description for the embedded case.
In the matrix regularization of string and M theories, the matrices 
$X^{\mu}$ are treated as the dynamical degrees of freedom and 
considered to realize a matrix regularization of the worldsheet of strings 
or the transverse worldvolume of membranes.

In this paper, we will consider the inverse problem 
of the construction of the matrix regularization. 
Namely, we start from a large-$N$ sequence of (bounded) 
matrices $\{X^{\mu}\}_N$, which may or may not 
realize a matrix regularization of a certain manifold, 
and consider how a classical space ${\cal M}$ can be 
associated with the matrix sequence.
This kind of problem is relevant for 
some physically interesting situations, such as the emergent 
geometry in the gauge/gravity correspondence and the 
matrix realization of membranes.
For this problem, (\ref{def of MR}) is not very 
useful since at the beginning
we do not know what the space ${\cal M}$ should be.
Instead, we define coherent states 
for a given sequence $\{X^{\mu}\}_N$
as a generalization of the known cases
and then define the classical space ${\cal M}$ 
as a set of all the coherent states.

For the cases where the classical space defined in this way 
forms a smooth manifold, 
we also express geometric objects on the classical space 
such as the metric, Levi-Civita connection, curvature and Poisson tensor,
in terms of the matrix elements.
If we consider a completely generic configuration of matrices, 
the classical space of our definition 
could be nonmanifold or an empty set. 
So in this sense, our result on the geometric objects 
cannot be applied to a generic matrix configuration
whose classical geometries are not smooth manifolds. 
However, as will be discussed in the last section, our method 
is still valid for a small perturbation around matrices which have 
smooth classical geometry.

The organization of this paper is as follows.
In Sec. \ref{Classical geometry of finite size matrices}, 
we introduce the notion of coherent states for a general matrix geometry and
define a classical space as a set of all coherent states. 
In Sec. \ref{Geometric objects}, 
we express the metric, connection, curvature and Poisson tensor on the 
classical space in terms of matrix elements.
In Sec. \ref{Examples}, 
we show some examples of our construction.
In Sec. \ref{Infinite dimensional matrices},
we consider the case of infinite-dimensional matrices.
In Sec. \ref{Summary and discussion},
we summarize our result and 
discuss possible applications.

\section{Classical geometry of finite size matrices}
\label{Classical geometry of finite size matrices}

In this section, we consider the classical (commutative) limit of finite 
size matrices. We denote an index set by $I := \{a_n | n\in {\bf N} \} $, 
where $\{ a_n \}$ is a strictly monotonically increasing sequence 
of natural numbers.
We assume that we are given a family of $D$ matrices, 
\begin{align}
\{ (X_1^{(N)},X_2^{(N)}, \cdots, X_D^{(N)} ) | N \in I \},
\end{align}
where $X_{\mu}^{(N)} \; (\mu =1,2,\cdots, D)$  are $N \times N$ Hermitian matrices.
We consider $ X_{\mu}^{(N)} $
as regularized embedding functions with the 
target space ${\bf R}^D$, which possibly realize 
a matrix regularization of a certain manifold. 
We raise and lower the $D$-dimensional indices
by using the Kronecker delta (the flat metric on ${\bf R}^D$),
so that we do not distinguish upper and lower indices in the following.

Let us define the position and its standard deviation
in ${\bf R}^D$ for each state\footnote{
In the following, we omit the superscript of $(N)$
and make the $N$-dependence implicit to avoid complexity. 
}.
Let ${\cal H}$ be the $N$-dimensional Hilbert space on which 
the $D$ matrices act and ${\cal H}^*$ be a subset of 
${\cal H}$ which consists of normalized state vectors:
\begin{align}
{\cal H}^* = \{ |\alpha \rangle \in {\cal H} | \langle \alpha | 
\alpha \rangle =1 \}.
\end{align}
For each $ |\alpha \rangle \in {\cal H}^*$, we define the 
position of $ |\alpha \rangle $ 
in ${\bf R}^D$ by
\begin{align}
x_{\mu}(|\alpha \rangle ) = 
\langle \alpha | X_{\mu} | \alpha  \rangle   \;\; (\mu =1,2,\cdots, D)
\end{align}
and the standard deviations by
\begin{align}
\sigma^2_{\mu} (| \alpha \rangle )
&=\langle \alpha | X_{\mu}^2 | \alpha \rangle 
-
\langle \alpha | X_{\mu} |\alpha \rangle^2
 \;\; (\mu =1,2,\cdots, D),
\nonumber\\
\sigma^2 (| \alpha \rangle )& = \sum_{\mu=1}^{D}
\sigma^2_{\mu} (| \alpha \rangle ).
\end{align}

Next, we introduce the notion of the coherent states.
We recall that the canonical coherent states can be defined as the 
ground states of the Hamiltonians,
$H(p_0,q_0) =\frac{1}{2}(\hat{p}-p_0)^2+\frac{1}{2}(\hat{q}-q_0)^2 $,
where $p_0$ and $q_0$ are real parameters 
(see Appendix \ref{Canonical coherent states}).
We generalize this definition.
We first introduce the ``Hamiltonian''
\begin{align}
H(y) =\frac{1}{2}\sum_{\mu=1}^D(X_{\mu}-y_{\mu})^2.
\label{hamiltonian}
\end{align}
This is an $N\times N $ Hermitian matrix defined for each point 
$y \in { \bf R }^D$.
The $N\times N$ identity matrix is omitted in the term of $y_{\mu}$.
We denote the $n$th eigenstate and the eigenvalue 
of $H(y)$ by $|n, y \rangle $ and $E_n(y)$:
\begin{align}
H(y)|n, y \rangle = E_n (y)|n, y \rangle \; \;\; ( n= 0,1,\cdots, N-1),
\end{align}
where we assume that the eigenvalues are ordered as 
$E_0(y) \leq E_1(y) 
\leq \cdots \leq E_{N-1}(y)$ and 
the states are normalized as
$\langle n,y |m,y \rangle = \delta_{mn}$.
Our definition of coherent states is the following.
We call $|0, y \rangle $ a coherent state at $y $ 
if and only if it satisfies\footnote{Note that strictly speaking 
the coherent state defined here is not a single state vector
but a set of the ground states which satisfy (\ref{lambda0 goes to 0}). }
\begin{align}
\lim_{N \rightarrow \infty} 
E_0 (y) = 0.
\label{lambda0 goes to 0}
\end{align}

Our definition of coherent states can be understood as follows.
Suppose that the coherent state exists at a point $y \in {\bf R}^D$. 
Since the ground state energy can be written as
\begin{align}
E_0 (y) = \frac{1}{2}\sigma^2 (|0,y\rangle )
+\frac{1}{2}\sum_{\mu =1}^{D}(x_{\mu}(|0,y\rangle)-y_{\mu})^2,
\end{align}
Eq. (\ref{lambda0 goes to 0}) implies that 
$\sigma (|0,y\rangle ) \rightarrow 0$ and 
$x_{\mu}(|0,y\rangle) \rightarrow y_{\mu}$.
Thus, it follows that 
there exists a state whose wave packet 
is centered at $y$ and shrinks to the point in the large-$N$ limit.
Conversely, suppose that there exists a state 
$| \alpha \rangle \in {\cal H}^*$
which satisfies $\sigma (|\alpha \rangle ) \rightarrow 0$ and
$x_{\mu}(| \alpha \rangle) \rightarrow y_{\mu}$.
Then, 
$\langle \alpha | H(y)| \alpha \rangle $ goes to zero in the 
large-$N$ limit.
Since $E_0(y)$ can be written as 
\begin{align}
E_0(y) = \min_{ |\psi\rangle \in {\cal H}^* }
 \langle \psi |H(y)|\psi \rangle ,
\end{align}
it also goes to zero in the large-$N$ limit; namely,
a coherent state exists at $y$. Therefore, for each $y \in {\bf R}^D$, 
if and only if a coherent state exists, 
there exists a wave packet which shrinks to a point in the large-$N$ limit.

It is natural to define the classical space, which we denote by ${\cal M}$, as 
a subspace of ${\bf R}^D$ on which there exist the shrinking wave functions.
In terms of the coherent states, 
this is equivalent to the subspace of ${\bf R}^D$ on which there exist 
the coherent states.
Let us define a function $f :{\bf R}^D \rightarrow {\bf R}_{+ }$ as
\begin{align}
f(y) = \lim_{N\rightarrow \infty}E_0(y).
\label{def of f}
\end{align}
Then, we can write ${\cal M}$ as
\begin{align}
{\cal M} =\{y\in  {\bf R}^D | f(y)=0  \}.
\label{def of M}
\end{align}
Note that one can compute $f(y)$ from the given matrices in principle.
So the expression (\ref{def of M}) provides a relation between 
the classical space and the matrix configurations.

Let us comment on some properties of the function $f(y)$.
Suppose that $E_0 (y) $ is a smooth function at a point
$y \in {\bf R}^D $, we can 
expand $E_0(y+\epsilon )$ in a Taylor series as
\begin{align}
E_0(y+\epsilon ) =
E_0(y) + \epsilon^{\mu} \partial_{\mu} E_{0 }(y) 
+ \frac{1}{2}\epsilon^{\mu}\epsilon^{\nu}
\partial_{\mu}\partial_{\nu} E_{0}(y) 
+\frac{1}{6}\epsilon^{\mu}\epsilon^{\nu}\epsilon^{\rho }
\partial_{\mu}\partial_{\nu}\partial_{\rho } E_{0}(y)
+ \cdots .
\label{Taylor}
\end{align}
Since $E_0(y+\epsilon )$ is the lowest eigenvalue of 
\begin{align}
H(y+\epsilon )=H(y)+\epsilon_{\mu}(y^{\mu}-X^{\mu}) 
+\frac{1}{2}(\epsilon^{\mu})^2,
\end{align}
we can compute the coefficients in (\ref{Taylor})
based on the perturbation theory as\footnote{Here, we 
assume that the Hamiltonian is nondegenerate, but we can 
also treat degenerate cases in a similar way. }
\begin{align}
\partial_{\mu} E_{0 }(y) & =  y_{\mu}- x_{\mu}( | 0, y \rangle ) , 
\nonumber\\
\partial_{\mu}\partial_{\nu} E_{0}(y) & = \delta_{\mu \nu }
- 2 \sum_{n =1}^{N-1}
{\rm Re} \frac{\langle 0, y | X_{\mu}| n,y \rangle 
 \langle n, y | X_{\nu}| 0,y \rangle }{E_n(y) -E_0 (y)},
\nonumber\\
\partial_{\mu}\partial_{\nu}\partial_{\rho} 
E_{0}(y) & =
\sum_{m \neq 0} \sum_{n \neq 0}
\frac{\langle 0, y| X_{\mu} |m, y \rangle
\langle m, y| X_{\nu} |n, y \rangle
\langle n, y| X_{\rho} |0, y \rangle }
{(E_m (y)-E_0(y))(E_n (y)-E_0(y))}
\nonumber\\
& \;\;\;\;\;\; 
- \langle 0, y | X_{\mu}|0, y \rangle 
\sum_{n \neq 0 } \frac{ \langle 0,y |X_{\nu} | n, y \rangle 
\langle n,y |X_{\rho} | 0, y \rangle }
{(E_n (y)- E_0 (y))^2} + \cdots ,
\label{lambda mu nu}
\end{align}
where $\cdots $ represents the sum over all permutations 
of the indices $\mu, \nu, \rho$.
We can also expand the function $f(y)$ as
\begin{align}
f(y+\epsilon ) =
f(y) + \epsilon^{\mu} \partial_{\mu} f(y) 
+ \frac{1}{2}\epsilon^{\mu}\epsilon^{\nu}
\partial_{\mu}\partial_{\nu} f(y) 
+\frac{1}{6}\epsilon^{\mu}\epsilon^{\nu}\epsilon^{\rho }
\partial_{\mu}\partial_{\nu}\partial_{\rho } f(y)
+ \cdots,
\label{expansion of f}
\end{align}
where $\partial_{\mu}f(y)$, $\partial_{\mu}\partial_{\nu}f(y)$ and 
$\partial_{\mu}\partial_{\nu}\partial_{\rho} f(y)$ are given by 
the large-$N$ limits of (\ref{lambda mu nu}).
Thus, we can compute the derivatives of $f(y)$ as
the corrections in the perturbation theory.
This property plays an important role, when we define 
geometric objects in terms of the matrix elements in the next section.
Note also that since $x^{\mu}(|0, y \rangle )$ goes to $ y^{\mu} $ 
for $y \in {\cal M}$ as discussed above, the single derivative of $f(y)$
is vanishing on ${\cal M}$:
\begin{align}
\partial_{\mu }f(y) =0  \;\; (\mu=1,2, \cdots, D, \;\; y \in {\cal M}).
\label{partial f is 0}
\end{align}
Hence, $f(y)$ is a quadratic function on a neighborhood of ${\cal M}$.




The coherent state at $y $ is an 
approximate simultaneous eigenstate of the $D$ matrices, 
where the eigenvalues are given by 
the components of $y $. 
In fact, it follows from (\ref{lambda0 goes to 0}) 
and the Cauchy-Schwarz inequality
that, for any 
vector $|\psi \rangle \in {\cal H}^*$, 
\begin{align}
\lim_{N \rightarrow \infty} \langle \psi | X^{\mu}-y^{\mu} |0,y \rangle 
= 0   \;\;\; (\mu=1,2, \cdots, D, \;\; y \in {\cal M}).
\label{cauchy}
\end{align}
From (\ref{cauchy}), one can also show the following equations
for $y\in {\cal M}$ and for any state vector $| \psi \rangle \in {\cal H}^*  $:
\begin{align}
&\lim_{N\rightarrow \infty }\langle \psi | A(X-y) | 0, y \rangle =0, 
\label{identity 0}
\\
&\lim_{N\rightarrow \infty } 
\left(
\langle \psi | A(X) | 0, y \rangle 
-
\langle \psi |0, y \rangle \langle 0, y |
 A(X) | 0, y \rangle 
\right) =0,
\label{identity of coh st 1}
\\
&\lim_{N\rightarrow \infty }
\langle \psi | [A(X), B(X)] | 0, y \rangle = 0,
\label{identity of coh st 2}
\end{align} 
where $A(X)$ and $B(X)$ are arbitrary 
polynomials of $X^{\mu}$ with finite ($N$-independent) 
degrees and coefficients. 
Here, we have assumed that 
$X^{\mu}$ are bounded in the large-$N$ limit.

\section{Geometric objects}
\label{Geometric objects}
In the following, we assume that 
there exists a neighborhood of ${\cal M}$ on which 
$E_0(y)$ is smooth\footnote{In order for
the curvature tensors to be defined, $E_0(y)$ needs to be at least 
3 times differentiable on the neighborhood.} 
for any sufficiently large $N$. 
In this case,
the classical space ${\cal M}$ is a smooth submanifold of ${\bf R}^D$.
In this section, under this assumption, 
we define various geometric objects such as 
the metric, Levi-Civita connection, curvature, and Poisson tensor 
on ${\cal M}$, in terms of the $D$ matrices $X^{\mu} $.
In the following, we assume for simplicity  
that the Hamiltonian (\ref{hamiltonian}) is nondegenerate 
on the neighborhood of ${\cal M}$, but the generalization 
to degenerate cases is straightforward.

\subsection{Metric}
We consider a real symmetric $D \times D$ matrix
\begin{align}
g_{\mu \nu}(y) =
2 \lim_{N\rightarrow \infty} 
\sum_{n=1}^{N-1}
 {\rm Re} \frac{\langle 0,y|X_{\mu}|n,y  \rangle
\langle n,y |X_{\nu}| 0,y \rangle}
{E_n(y)-E_0(y)}.
\label{def of G}
\end{align}
Note that this can also be written as 
\begin{align}
g_{\mu \nu}(y) = \delta_{\mu \nu} - \partial_{\mu }\partial_{\nu }f(y).
\label{g=delta-f}
\end{align}
In the following, we show that 
the matrix $g_{\mu \nu}(y)$ is
a metric on ${\cal M}$.

Let us consider the expansion (\ref{expansion of f}) in which
$|\epsilon | = \sqrt{\delta_{\mu\nu} 
\epsilon^{\mu}\epsilon^{\nu}}$ 
is much smaller than 
the typical scale of ${\cal M}$.
If both $y $ and $y+\epsilon $ are contained in ${\cal M}$, 
we can regard $\epsilon $ as a tangent vector on ${\cal M}$, 
and we denote $\epsilon = \epsilon_{\parallel }$.
In this case, since $f(y)=f(y+\epsilon_{\parallel } )=0 $, we have
\begin{align}
g_{\mu \nu}(y) \epsilon^{\mu}_{\parallel}
\epsilon^{\nu}_{\parallel} = |\epsilon_{\parallel} |^2.
\label{parallel}
\end{align}
On the other hand, let us consider the case in which
$\epsilon= \epsilon_{\perp} $ 
is a normal vector at $y \in {\cal M}$.
In this case, as shown in Appendix \ref{Derivation 1}, 
$f(y+\epsilon_{\perp} )$ is given by the distance between 
$y$ and $y+ \epsilon_{\perp} $ as 
\begin{align}
f(y+\epsilon_{\perp} ) = \frac{1}{2}| \epsilon_{\perp} |^2.
\label{distance squared}
\end{align}
By comparing this to Eq. (\ref{expansion of f}) with 
$\epsilon= \epsilon_{\perp} $, we obtain
\begin{align}
g_{\mu \nu}(y) \epsilon^{\mu}_{\perp}
\epsilon^{\nu}_{\perp } = 0.
\label{perp}
\end{align}
Equations (\ref{parallel}) and (\ref{perp}) show that 
$g_{\mu \nu}(y)$ is a projection to the tangent space $T{\cal M}_y$
at $y\in {\cal M}$.
Thus, it satisfies
\begin{align}
g_{\mu \nu}(y)g^{\nu}{}_{ \rho}(y) = g_{\mu \rho }(y)  \;\; (y \in {\cal M}).
\label{projection property}
\end{align}
Here, we again emphasize that the index $\nu $ is raised using 
the Kronecker delta, so that $g^{\nu}{}_{ \rho}=g_{\nu \rho}$.
Since this implies that 
$g_{\mu \nu}(y)$ is positive and nondegenerate on ${\cal M}$, 
it gives a metric on ${\cal M}$. The line element on ${\cal M}$ can be
written as 
\begin{align}
ds^2 = g_{\mu \nu}(y) dy^{\mu}dy^{\nu}.
\label{induced metric}
\end{align}
It is interesting that the positivity of the metric comes from 
the negativity of the second-order correction to the ground state energy 
in the perturbation theory of quantum mechanics.

Note that, for any $y \in {\cal M}$, the dimension of ${\cal M}$ is 
given by the trace of $g_{\mu \nu}(y)$\footnote{If ${\cal M}$ consists
of some disconnected components, the trace of $g_{\mu \nu}(y)$ 
gives the dimension of 
the component which contains $y$.}, 
\begin{align}
{\rm dim }{\cal M}  = \sum_{\mu =1}^D g_{\mu \mu}(y).
\end{align}
This expression and (\ref{def of G}) relate the 
dimension of ${\cal M}$
with the matrix configurations.

\subsection{Levi-Civita connection}
Let $A^{\mu} (y)$ and $B^{\mu}(y)$ be tangent vector fields which 
satisfy $g^{\mu }{}_{ \nu }(y) A^{\nu} (y) = A^{\mu} (y)$ and 
$g^{\mu }{}_{ \nu }(y) B^{\nu} (y) = B^{\mu} (y)$
for $y \in {\cal M}$.
We define the covariant derivative on ${\cal M}$ by 
\begin{align}
(\nabla_B A)^{\mu} = B^{\nu}(\partial_{\nu }A^{\mu }
+\Gamma^{\mu}_{\nu \rho}A^{\rho} ).
\label{CovD}
\end{align}
The connection $\Gamma^{\mu}_{\nu \rho} $ is
chosen such that the image of the covariant derivative 
(\ref{CovD}) is again a tangent vector, namely, 
\begin{align}
g^{\mu }{}_{ \nu }(\nabla_B A)^{\nu}=
(\nabla_B A)^{\mu}.
\label{def of connection}
\end{align}
By using (\ref{g=delta-f}) and 
\begin{align}
B^{\rho }(y) \left\{ \partial_{\rho }g_{\mu \sigma }(y)
-g_{\mu \nu}(y)\partial_{\rho }g^{\nu }{}_{ \sigma }(y)
-g_{\nu \sigma }(y)\partial_{\rho }g^{\nu}{}_{\mu }(y)
\right\} =0
\;\; (y \in {\cal M}),
\label{derivative of projection property}
\end{align}
which is obtained by differentiating 
(\ref{projection property}), we find 
a simple solution to (\ref{def of connection}) as
\begin{align}
\Gamma^{\mu}_{\nu \rho} =
(\partial^{\sigma }\partial^{\mu }f )
( \partial_{\sigma }
\partial_{\nu }\partial_{\rho }f ).
\label{our choice of connection}
\end{align}
We also find that $(\nabla_B g)_{\mu \nu} =0$. 
This means that   (\ref{our choice of connection})
is the Levi-Civita connection associated with
 the metric $g_{\mu \nu}$.
The expression (\ref{our choice of connection})
 together with (\ref{lambda mu nu}) 
relates the connection  and the 
matrix elements of $X^{\mu} $.

\subsection{Curvature}
\label{Curvature}
Let $A,B,C $ be tangent vector fields on ${\cal M}$. 
The curvature tensor is defined by
\begin{align}
R(A,B)C = [\nabla_A, \nabla_B ]C -\nabla_{[A,B]}C,
\end{align}
where $[A, B] $ represents the Lie bracket of the vector fields, 
\begin{align}
[A, B]^{\mu}= A^{\nu}\partial_{\nu}B^{\mu}- B^{\nu}\partial_{\nu}A^{\mu}.
\end{align}
Note that this is also a tangent vector satisfying 
$g_{\mu \nu}[A, B]^{\nu}=[A, B]^{\mu}$.
In terms of components, we can write the curvature as
\begin{align}
(R(A,B)C)^{\mu}
&= A^{\nu }B^{\rho}C^{\sigma}
R^{\mu}{}_{\sigma \nu \rho}
\nonumber\\
&=A^{\nu }B^{\rho}C^{\sigma}
\left\{
(\partial^{\mu}\partial_{\nu}\partial_{\lambda}f)
(\partial_{\rho}\partial_{\sigma}\partial^{\lambda}f)
-
(\partial^{\mu}\partial_{\rho}\partial_{\lambda}f)
(\partial_{\nu}\partial_{\sigma}\partial^{\lambda}f)
\right\}.
\end{align}
By contracting the indices of $R^{\mu}{}_{\sigma \nu \rho}(y)$ 
using $g_{\mu \nu}(y)$, we can also 
write the Ricci tensor or the Ricci scalar in terms of 
the derivatives of $f(y)$.
These expressions and (\ref{lambda mu nu}) give relations
between the curvature tensors and the matrices.

\subsection{Poisson tensor}
We consider a real antisymmetric $D \times D$ matrix on ${\cal M}$ defined by 
\begin{align}
W^{\mu \nu }(y)= \lim_{N\rightarrow \infty} c \langle 0, y |[X^{\mu}, X^{\nu }] |0,y \rangle ,
\label{poisson}
\end{align}
where $c $ is a pure imaginary $N$-dependent normalization constant.
The constant $c$ is chosen so that 
$c \langle 0, y |[X^{\mu}, X^{\nu }] |0,y \rangle$ 
becomes ${\cal O}(N^0)$ in the large-$N$ limit 
(see the next section for concrete examples).
$W^{\mu \nu }(y) $ is a tangent bivector on ${\cal M}$ and 
satisfies the Jacobi identity as we will show below. 
This means that $W^{\mu \nu }(y) $ is a Poisson tensor on 
${\cal M}$.

We first show that $W^{\mu \nu }(y) $ is a tangent bivector, 
namely, it satisfies $g_{\mu \rho }(y)W^{\rho \nu }(y) =W^{\mu \nu }(y) $
for $y \in {\cal M}$.
In order to prove this, the following relation is useful:
\begin{align}
c \langle 0,y | [X^{\mu }, X^{\nu }] | 0,y \rangle 
\langle 0,y | X_{\nu} | n, y \rangle  
\sim c (E_{n}(y)-E_{0}(y))
\langle 0,y | X^{\mu} | n, y \rangle ,
\label{useful}
\end{align}
where $\sim $ stands for an equality for the leading-order terms in the 
large-$N$ limit.
This can be shown as follows. 
By using (\ref{identity of coh st 1}) and 
(\ref{identity of coh st 2}), we obtain
\begin{align}
c \langle 0,y | [X^{\mu }, X^{\nu }] | 0,y \rangle 
\langle 0,y | X_{\nu} | n, y \rangle  
& \sim 
\frac{c}{2} \langle 0,y |\{ X^{\nu}-y^{\nu}, [X^{\mu }, X_{\nu }-y_{\nu}]\} 
| n,y \rangle .
\end{align} 
Then noticing that 
$\{ X^{\nu}-y^{\nu}, [X^{\mu }, X_{\nu }-y_{\nu}]\} 
= 2[X^{\mu}, H(y)]$, 
we obtain (\ref{useful}).
Now, let us calculate 
\begin{align}
g^{\mu }{}_{ \rho }(y)W^{\rho \nu }(y) 
= \lim_{N \rightarrow \infty } 
\left[ 
2 {\rm Re} 
\left \{
c \sum_{n \neq 0} 
\frac{\langle 0, y | X^{\mu}|n,y \rangle 
\langle n, y | X^{\rho }| 0,y \rangle }
{E_n (y)- E_0 (y)} \langle 0, y | [X_{\rho}, X^{\nu}] |0,y \rangle
\right\}
\right] .
\end{align}
By using (\ref{useful}) and the completeness relation 
$\sum_{n=0}^{N-1} |n, y \rangle \langle n,y | =1 $, we obtain 
\begin{align}
g^{\mu }{}_{ \rho }(y)W^{\rho \nu }(y) = 
\lim_{N \rightarrow \infty } 
\left[ 
2 {\rm Re} 
\left \{ 
c \langle 0, y | X^{\mu }X^{\nu } | 0,y \rangle 
-c \langle 0, y | X^{\mu }| 0,y \rangle 
\langle 0,y | X^{\nu } | 0,y \rangle 
\right\}
\right].
\end{align}
Since we have assumed that $c $ is pure imaginary and 
$X^{\mu}$ are Hermitian, 
the second term is zero, while the 
first term is equal to $W^{\mu \nu}(y)$.
Thus, we have shown that $W^{\mu \nu}(y)$ is a 
tangent bivector on ${\cal M}$.
The Jacobi identity of the Poisson bracket 
can be shown in a similar way. We 
leave the proof to Appendix \ref{Jacobi identity}.

\section{Examples}
\label{Examples}
In this section, we consider some examples. 

\subsection{Fuzzy sphere}
Our first example is the fuzzy sphere, which is given by 
\begin{align}
X_{\mu} = \frac{2}{\sqrt{N^2 -1}}L_{\mu} \;\; (\mu =1,2,3),
\label{fuzzy sphere config}
\end{align}
where $L_{\mu}$ are the $N$-dimensional representation matrices of 
the $SO(3)$ generators satisfying $[L_{\mu}, L_{\nu}]= 
i\epsilon_{\mu \nu \rho}L_{\rho}$ and 
$\sum_{\mu =1}^3  L^2_{\mu} =\frac{N^2-1}{4} {\bf 1}_{N \times N}$. 
The normalization in (\ref{fuzzy sphere config}) is chosen in such a way that 
the radius of the sphere becomes one: $X^{\mu}X_{\mu}= {\bf 1}_{N \times N}$.
The corresponding Hamiltonian (\ref{hamiltonian}) is given by 
\begin{align}
H(y)= \frac{1+|y|^2}{2}- \frac{2L_{\mu} y^{\mu} }{\sqrt{N^2-1}}.
\end{align}
Here, we notice that for any $SO(3)$ rotation,
$L_{\mu} \rightarrow \Lambda_{\mu}{}^{\nu}L_{\nu}$,
there exists a unitary transformation which reproduces
this rotation:
\begin{align}
RL_{\mu}R^{\dagger }= \Lambda_{\mu}{}^{\nu}L_{\nu}.
\end{align}
See Appendix \ref{Bloch coherent states} for this transformation.
Hence it suffices to consider the case where 
$y= (y^1,y^2,y^3)= (0,0,r)$ with $r=|y| \geq 0$.
Then, the Hamiltonian reduces to
\begin{align}
H(y)= \frac{1+|y|^2}{2} - \frac{2L_{3} |y| }{\sqrt{N^2-1}}.
\label{reduced Ham fs}
\end{align}
The ground state of this Hamiltonian is given by the 
highest eigenstate of $L_3 $ with the eigenvalue $(N-1)/2$. 
Thus, we find that the function $f(y)$ is given by 
\begin{align}
f(y)=\frac{1}{2}(1-|y|)^2.
\end{align}
Therefore, the classical space ${\cal M}$, which is given as a set of 
zeros of $f(y)$, is indeed a sphere with the unit radius.
Here, we remark that $E_0(y)$ is 
a smooth function for any point $y$ apart from the origin,
so that the smoothness condition assumed in the previous section is 
satisfied.

We can also compute the metric either from 
(\ref{def of G}) or (\ref{g=delta-f}).
The result is given by
\begin{align}
g_{\mu \nu }(y) =\frac{1}{|y|}
\left( \delta_{\mu \nu}- \frac{y_{\mu} y_{\nu}}{|y|^2} \right). 
\end{align}
For $y \in {\cal M}$, this is indeed a projection to the 
tangent space and 
(\ref{induced metric}) gives the standard line element
on $S^2$. It is an easy exercise to compute the curvature tensors from 
the definition in Sec. \ref{Curvature}
 and check that 
the Ricci scalar is equal to $2$ and agrees with 
the known value for the unit sphere. 

The Poisson tensor (\ref{poisson}) is given by 
\begin{align}
W^{\mu \nu }(y)= i\epsilon^{\mu \nu \rho} 
\lim_{N\rightarrow \infty} \frac{2c}{\sqrt{N^2-1}} 
\langle 0, y |X_{\rho} |0,y \rangle.
\end{align}
If we put $c=\frac{-i }{2 }\sqrt{N^2 -1}  $, we obtain
\begin{align}
W^{\mu \nu }(y) = \epsilon^{\mu \nu \rho} y_{\rho},
\end{align}
where we have used (\ref{cauchy}).
This is the standard Poisson tensor on $S^2$ 
embedded in ${\bf R}^3$.

\subsection{Fuzzy torus}
The fuzzy torus is defined in terms of 
 the algebra of two unitary matrices $U$ and $V$,
\begin{align}
VU =e^{i\theta }UV,
\label{alg for FT}
\end{align}
where  $\theta=2\pi /N$.
The matrices $U$ and $V$ can be represented by the
clock and shift matrices as
\begin{align}
&U_{mn} = \delta_{mn}e^{in\theta } , 
\nonumber\\
&V_{mn} = \delta_{m+1 n},
\label{clock shift}
\end{align}
where $m,n =0, 1, \cdots, N-1$ and 
$\delta_{mn}$ is the cyclic 
Kronecker delta which satisfies
$\delta_{Nn}\equiv \delta_{0n}$.
By introducing four Hermitian matrices $X^{\mu}\; (\mu =1,2,3,4)$ as 
the real and imaginary parts of $U$ and $V$ as
\begin{align}
U=X^1+iX^2, \;\;\; V=X^3+iX^4, 
\end{align}
we can regard the fuzzy torus as the noncommutative 
 Clifford torus embedded in ${\bf R}^4$.
Note that $X^1$ and $X^2$ commute with each other and satisfy 
$(X^1)^2+(X^2)^2=  {\bf 1}_{N \times N} $ because of 
the unitarity condition of $U$, and so do $X^3$ and $X^4$
because of 
the unitarity condition of $V$.
The Hamiltonian (\ref{hamiltonian}) is given by
\begin{align}
H(y)= \frac{1}{2}(U-z)(U^{\dagger}-\bar{z })+
\frac{1}{2}(V-w)(V^{\dagger}-\bar{w }),
\label{H for FT}
\end{align}
where $z=y^1+iy^2$ and $w=y^3+iy^4$.
As shown in Appendix
\ref{Classical geometry for fuzzy torus}, 
in this case the function $f(y)$ is given 
by 
\begin{align}
f(y)=\frac{1}{2}(1-|z|)^2 +\frac{1}{2}(1-|w|)^2.
\label{F for FT}
\end{align}
Hence, the classical space ${\cal M}$ is indeed the Clifford
 torus given by $|z|=|w|=1$.

We can also obtain the metric from 
(\ref{g=delta-f}) as
\begin{align}
&g_{a b }(y) =\frac{1}{|z|}
\left( \delta_{a b}- \frac{y_{a} y_{b}}{|z|^2}\right),
\;\;\;  g_{a \alpha }(y)=0,
\;\; \;
g_{\alpha \beta }(y)= \frac{1}{|w|}
\left(
\delta_{\alpha \beta}- \frac{y_{\alpha } y_{\beta}}{|w|^2}
\right), 
\end{align}
where $a, b =1,2$ and $\alpha, \beta =3,4$.
We can see that $g_{\mu \nu}$ is a projection operator
 for $y \in {\cal M}$ and
gives the standard metric of the Clifford torus.

In order to compute the Poisson tensor, 
we rewrite the algebra (\ref{alg for FT}) using $X^{\mu } $ as 
\begin{align}
&[X_1\pm iX_2 , X_3 \pm iX_4 ] \sim i \theta (X_1\pm iX_2)(X_3 \pm iX_4),
\nonumber\\
&[X_1\pm iX_2 , X_3 \mp iX_4 ] \sim - i \theta (X_1\pm iX_2)(X_3 \mp iX_4),
\end{align}
where we have neglected higher-order terms in $\theta$.
By putting $c = i/\theta $ in (\ref{poisson}), 
we obtain the Poisson tensor on the Clifford torus:
\begin{align}
  W^{\mu \nu}(y) = \left(
    \begin{array}{cccc}
      0 & 0 & y^2y^4 & -y^2y^3 \\
      0 & 0 & -y^1y^4 & y^1y^3 \\
       -y^2y^4 & y^1y^4 & 0 & 0  \\
       y^2y^3 & -y^1y^3 & 0 & 0 \\
    \end{array}
  \right).
\end{align}

\section{Infinite-dimensional matrices}
\label{Infinite dimensional matrices}

As far as the Hamiltonian (\ref{hamiltonian}) has a 
discrete spectrum, it will be possible  to 
apply the same formulation to matrices with an infinite size. 
In this case, we assume that we are given a one-parameter family 
of $D$ operators $ X_{\mu}^{ (\hbar )} $ 
acting on an infinite-dimensional Hilbert space ${\cal H}_{\hbar}$, 
where $\hbar $ is a non-negative parameter.
We replace the large-$N$ limit considered in the previous sections 
with the limit of $\hbar \rightarrow 0 $ and 
the finite sums such as $\sum_{n=0}^{N-1}$ with the infinite sums 
$\sum_{n=0}^{\infty}$ appropriately.
In this section, by considering a generic example, 
we will demonstrate that our formulation also works for 
infinite-dimensional matrices.

In this section, we work with the notation as follows.
We decompose the indices $\mu, \nu, \rho $, which 
run from $1$ to $D$, to two indices as $\mu \rightarrow \{A, I\}$.
The indices $A, B, C, \cdots$ run from 
$1$ to 4, while $I, J, K, \cdots $ run from $5$ to $D$.
We further decompose $A \rightarrow \{a, \alpha \}$, 
so that $a, b, \cdots $ run from $1$ to $2$ and
$\alpha, \beta, \cdots $ run from $3$ to $4$.

We consider a four dimensional noncommutative 
plane with some fluctuations embedded in ${\bf R}^D$,
which is given by 
\begin{align}
&X^A = \hat{q}^A +\tilde{X}^A  \; \; (A=1,2,3,4),
\nonumber\\ 
&X^I = \tilde{X}^I  \; \; (I=5,6,\cdots, D).
\label{config on Moyal}
\end{align}
Here $\hat{q}^A$ are the operators satisfying
\begin{align}
&[\hat{q}^a, \hat{q}^b] = i \epsilon^{ab} \theta, \;\;\; 
[\hat{q}^{\alpha }, \hat{q}^{\beta }] = i \epsilon^{\alpha \beta} \theta', \;\;\;
[\hat{q}^{a }, \hat{q}^{\alpha }] = 0, 
\end{align}
where $\epsilon^{ab}$ and 
$\epsilon^{\alpha \beta}$ are
$2\times 2 $ antisymmetric matrices with $\epsilon^{12}=1$ 
and $\epsilon^{34}=1$. 
$\theta$ and $\theta'$ 
are the noncommutative parameters which are assumed to vanish 
in the classical limit $\hbar \rightarrow 0 $.
We assume that the fluctuations $\{\tilde{X}^{\mu} \}$ are 
of the form
\begin{align}
\tilde{X}^{\mu }= 
\int \frac{d^4k}{(2\pi)^4}\int d^4x e^{-ik\cdot x} e^{ik \cdot \hat{q}}
\phi^{\mu}(x) \;\;\; (\mu =1,2, \cdots, D),
\end{align}
where $k\cdot x = \sum_{A=1}^4 k_Ax^A$ is the four-dimensional 
inner product and 
the Weyl symbols $\phi^{\mu}(x)$ are real functions
on ${\bf R}^4$. Here, we emphasize that 
$\phi^{\mu}(x)$ are just C-number valued functions
and they depend only on the four-dimensional coordinates $ x_A  $.
We assume that $|\phi^{\mu}(x)|$ are small, so that we can treat 
$\tilde{X}^{\mu}$
perturbatively. 
The configuration (\ref{config on Moyal}) 
is often considered in the context of the description of
gauge theories on the noncommutative plane
in terms of matrix models\cite{Aoki:1999vr}.

For the configuration (\ref{config on Moyal}),
the Hamiltonian (\ref{hamiltonian}) is given by 
\begin{align}
H(y)= H_0(y) + H_1(y) + H_2(y),
\end{align}
where $H_0(y), H_1(y)$ and $H_2(y)$ are defined by
\begin{align}
H_0(y)= \frac{1}{2} (\hat{q}_A -y_A)^2 +\frac{1}{2}y_I^2, 
\;\;\; 
H_1(y)= \frac{1}{2} \{ \hat{q}_A -y_A , \tilde{X}^A \} -\tilde{X}_Iy^I,
\;\;\;
H_2(y)= \frac{1}{2}\tilde{X}_{\mu}^2.
\label{Ham for infinite}
\end{align}
The subscripts of the Hamiltonians indicate the degrees of the perturbation.
The zeroth-order Hamiltonian $H_0(y)$ is just the 
two-dimensional harmonic oscillator. 
We introduce eigenstates of $H_0(y)$ by
\begin{align}
&H_0(y)|m,n,y \rangle = E^{(0)}_{m,n}|m,n,y \rangle, 
\nonumber\\
&E^{(0)}_{m,n} = \theta \left(m+ \frac{1}{2} \right)
+ \theta' \left(n+ \frac{1}{2} \right) +
\frac{1}{2}y_I^2.
\end{align}
See Appendix \ref{Derivation of equation} for 
explicit forms of the wave functions.
In the following, in order to
avoid the degeneration of $H_0(y)$,
we assume that $\theta$ and $\theta'$ are written as 
$\theta = a \hbar $ and $\theta' =b\hbar $, 
where $a$ and $b$ are rational and irrational 
constants, respectively.
Under this assumption, we can naively 
apply the formulas in the previous sections.

In Appendix \ref{Derivation of equation}, we compute the 
ground state energy of the full Hamiltonian $H(y)$ 
up to the second-order perturbation in $\tilde{X}_{\mu}$ 
and derive the following form of $f(y)$:
\begin{align}
f(y)= \frac{1}{2}
\left\{ y^I-\phi^I(y)+\phi^A(y)\partial_A \phi^I(y) \right\}
h_{IJ}(y)
\left\{ y^J-\phi^J(y)+\phi^B(y)\partial_B \phi^J(y) \right\} 
+ {\cal O}(\phi^3),
\label{f for moyal}
\end{align}
where 
\begin{align}
h_{IJ}(y)= \delta_{IJ}-\partial^A \phi_I(y)\partial_A \phi_J(y).
\end{align}
Hence, the classical space ${\cal M}$ is given by a four-dimensional 
surface in ${\bf R}^D$ defined by
\begin{align}
y_I = \phi_I(y)-\phi^A(y)\partial_A \phi_I(y) + {\cal O}(\phi^3)
\;\;\; (I=5,6,\cdots, D).
\label{cal M for NCplane}
\end{align}
When $\phi^{\mu}=0$, the above equation reduces to $y_I=0$ and 
${\cal M}$ is just the four-dimensional flat space parametrized by 
$\{y_A \}$. Equation (\ref{cal M for NCplane}) shows that
${\cal M}$ also fluctuates along the transverse directions
when $\phi^{\mu}$ are turned on.

From Eq. (\ref{def of G}) or (\ref{g=delta-f}), we can 
read off the metric on $ {\cal M}$ as 
\begin{align}
&g_{AB}(y)=\delta_{AB}-\partial_A \phi_I(y) \partial_B \phi^I(y)+ {\cal O}(\phi^3), 
\nonumber\\
&g_{AI}(y)=\partial_A \phi_I(y) -\partial_A(\phi^B(y) \partial_B \phi_I(y))+ {\cal O}(\phi^3), 
\nonumber\\
&g_{IJ}(y)=\partial^A\phi_I(y)\partial_A\phi_J(y)+ {\cal O}(\phi^3).
\label{metric components}
\end{align}
It is easy to check the projection property 
(\ref{projection property}) for this metric.
By substituting the metric (\ref{metric components}) 
into (\ref{induced metric}), 
we obtain the line element
\begin{align}
ds^2 = (\delta_{AB}+\partial_A \phi^I (y) \partial_B \phi_I(y))dy^A dy^B+ {\cal O}(\phi^3),
\end{align}
where we have used $dy^I=g_{IA}(y)dy^A$ obtained from 
(\ref{cal M for NCplane}).
By introducing new coordinates $\sigma^A$ 
by $y^A = \sigma^A +\phi^A(\sigma) $, 
we can rewrite the line element as 
\begin{align}
ds^2 = \big[
(\delta_{A}^C+\partial_A \phi^C (\sigma))
(\delta_{BC}+\partial_B \phi_C (\sigma)) 
+ \partial_A \phi_I (\sigma)\partial_B \phi^I (\sigma) 
\big]
d\sigma^A d\sigma^B+ {\cal O}(\phi^3).
\end{align}
This is nothing but the induced metric associated with 
the embedding function given by $X^A(\sigma)=\sigma^A + \phi^{A}(\sigma )$ and 
$X^I(\sigma)= \phi^{I}(\sigma )$.

Through a similar calculation, 
we can obtain the following form for the Poisson tensor,
\begin{align}
W^{AB}(y) = &B^{AB} 
+B^{AC}\partial_C \phi^B(y)
-B^{BC}\partial_C \phi^A(y)
+B^{CD}\partial_C \phi^A(y)\partial_D \phi^B(y)
\nonumber\\
& -B^{AD} \phi^C(y)\partial_C \partial_D \phi^B(y)
+B^{BD} \phi^C(y)\partial_C \partial_D \phi^A(y)+ {\cal O}(\phi^3),
\nonumber\\ 
W^{AI}(y) =&B^{AB}\partial_B \phi^I (y)
+B^{BC}\partial_B \phi^A (y)\partial_C \phi^I (y)
-B^{AD} \phi^C (y)\partial_C \partial_D \phi^I (y)+ {\cal O}(\phi^3),
\nonumber\\ 
W^{IJ}(y) =&B^{AB}\partial_A \phi^I (y) \partial_B \phi^J (y)+{\cal O}(\phi^3).
\label{Poisson components}
\end{align}
Here, we have defined a real antisymmetric matrix $B^{AB}$ by 
$[\hat{q}^A, \hat{q}^B] = i \hbar B^{AB} $ and we have 
chosen the constant $c$ in (\ref{poisson}) as $c = -i/\hbar $.
From (\ref{metric components}) and 
(\ref{Poisson components}),
one can easily check that $W^{\mu \nu}(y)$ is a tangent bivector and 
its Poisson bracket satisfies the Jacobi identity.

\section{Summary and discussion}
\label{Summary and discussion}

In this paper, we proposed a novel method of finding the classical limit
of matrix geometries.
We first introduced the notion of coherent states for general matrix geometry 
described by a large-$N$ sequence of
$D$ Hermitian matrices $(X^1, X^2, \cdots, X^{D})$.
We then defined the classical space ${\cal M}$ as a set of all 
coherent states. Assuming that the set of coherent states
forms a smooth manifold and so does ${\cal M}$,
we also found expressions for various geometric objects on ${\cal M}$ 
(metric, connection, curvature and Poisson tensor) in terms of 
the $D$ matrices.

The usage of our result on the geometric objects 
is limited to the cases 
where the classical space is a smooth manifold. 
Probably, if we consider a general matrix sequence, the corresponding classical space is most likely to be an empty set. 
In order to have a nontrivial space, there must be a subspace of the Hilbert 
space on which the $D$ matrices become mutually commuting with each other.
Though this is not the case for a general sequence, 
there are physically relevant situations where
such commuting matrices naturally arise. 
In matrix models of the Yang-Mills type (i.e. models 
with commutator interactions), typical values of 
the commutators might become very small 
in the strong coupling region, since one can always take a normalization 
such that the coupling constant appears only in front of the commutators.
As stated in \cite{Berenstein:2008eg}, 
this gives a reasonable mechanism for the emergence of space in matrix models.
Thus, in the strong coupling regime of matrix models, 
we can expect that nontrivial manifolds emerge and they can be visualized 
by using our method.

Even if the space is nonempty, 
we needed to assume the smoothness condition to define the geometric objects.
This condition is necessary to avoid some singular cases. 
The simplest singular example
can be found for some special cases of
diagonal matrices. Suppose that
all of the $D$ matrices are diagonal for any $N$.
In this case, ${\cal M}$ is just given by a discrete 
set of points corresponding to 
the positions of the eigenvalues in ${\bf R}^D$.
Let us consider a further special case where 
the eigenvalue distribution becomes dense and continuous 
in a certain fixed region of ${\bf R}^D$ in the large-$N$ limit, 
as the limit considered in the context of the large-$N$ 
reduction\cite{Eguchi:1982nm,Bhanot:1982sh,Parisi:1982gp,Gross:1982at}. 
In this limit,
${\cal M}$ becomes continuous manifold given by the
eigenvalue distribution in the large-$N$ limit.
However, near the large-$N$ limit, 
the ground state energy $E_0(y) $ is nondifferentiable 
almost everywhere on the dense region.
Hence the formulation in Sec. \ref{Geometric objects}
does not work.
This example suggests that the off-diagonal 
elements are essential to describe the metric and the other geometric objects.
Probably this kind of singularity should also be related with 
the topology change of membranes in M theory and 
we hope to find clear criteria 
to distinguish such singular configurations of matrices.

Let us make some comments on our results.
Firstly, our result can easily be applied to 
small perturbations around 
smooth configurations. 
For finite $N$ matrices, let us consider an expansion of the matrices, 
$X^{\mu} = \hat{X}^{\mu}+h^{\mu}(\hat{X})$. Here,
$\hat{X}^{\mu}$ are (a large-$N$ sequence of) matrices which 
satisfy the smoothness condition and have an associated classical manifold ${\cal M}$.
$h^{\mu}(\hat{X})$ are smooth polynomial functions which represent small 
fluctuations. In this case, by treating $h^{\mu}(\hat{X})$ as perturbation,
from (\ref{identity 0})-(\ref{identity of coh st 2}),
one can show that the classical space for $X^{\mu}$ is given by 
$\{ y\in {\bf R}^D | y^{\mu}=\hat{y}^{\mu}+
(\delta^{\mu}_{\nu}-\hat{g}^{\mu}{}_{\nu}(\hat{y}))
h^{\nu}(\hat{y}), \;\; \hat{y} \in {\cal M} \}$,
where $\hat{g}_{\mu \nu}$ is the metric of ${\cal M}$.
Thus, the perturbed geometry is again a smooth manifold.

Secondly, if the matrices are thought of as the transverse coordinates 
of D-branes, the classical space just corresponds to the 
classical shape of the D-branes.
For example, the fuzzy sphere formed by matrices of D0-branes 
can be interpreted as the D2-brane arising via the Myers effect
\cite{Myers:1999ps}.
In this phenomena, the net D2-brane charge is zero, while 
there is a nontrivial gauge flux on D2-brane, which induces a coupling 
to the R-R 1-form. This gauge field will be realized as 
the Berry connection with respect to the Hamiltonian (\ref{hamiltonian}) 
in our formulation. 
For example in the fuzzy sphere case, let us define 
$A_{\mu}=-i \langle \Omega | \partial_{\mu}|\Omega \rangle  $, where
$|\Omega \rangle$ is the Bloch coherent state defined in (\ref{def of bloch})
and $\mu = \theta, \phi$. Then
we can find that the gauge field coincides with 
the Dirac monopole configuration, which 
agrees with the gauge field induced on D2-branes \cite{Myers:1999ps}.
Although we have discussed local properties of the coherent states in this
paper, it will also be interesting to study global aspects in 
connection with the Berry phase of the Hamiltonian (\ref{hamiltonian}).

Now, let us discuss some possible applications of our result.
Since all the geometric objects defined in this paper 
are invariant under the $U(N)$ gauge transformation 
$X^{\mu}\rightarrow UX^{\mu }U^{\dagger} $, they
can be regarded as observables in matrix models.
These new observables will be useful in understanding 
geometric aspects of matrix models.
For example, they can be used to 
figure out the classical shape of 
D-branes in matrix models \cite{Azeyanagi:2009zf} 
and also to study the emergence of space-time 
in matrix models\cite{Steinacker:2007dq,Kim:2011cr}.

Our formulation also works for higher-dimensional gauge theories,
and we can define the similar observables in gauge theories,
which are useful in studying the gauge/gravity 
correspondence\cite{Maldacena:1997re}.
In the gauge/gravity correspondence, 
the background geometry of the string theory is expected 
to be emergent in the strong coupling limit 
of the corresponding gauge 
theory\cite{Berenstein:2005aa,Asano:2014eca,Asano:2014vba}.
Though field configurations on the gauge theory side 
are considered to encode the information of the background geometry, 
a general prescription to see this is not known yet. 
Our observables will offer a new method of translating
field configurations to the corresponding background geometry. 

Those observables will also be quite useful in
computer simulations of matrix models
\cite{Kim:2011cr,Catterall:2007fp,Anagnostopoulos:2007fw}. 
By using a computer, one can easily 
compute the observables from given matrix configurations 
and see the geometric properties.
Furthermore, although we considered the strict large-$N$ limit
when we defined the geometric objects, 
our formulation also works 
at finite $N$ as an approximation,
which is rather easily accessible from numerical 
simulations\footnote{ For example, the dimension of the fuzzy sphere 
(\ref{fuzzy sphere config}) with $N=50$ 
can be read off from the value of $g_{\mu\mu}$ at the minimum of 
the ground state energy. 
One can find $g_{\mu\mu}=2.04$, which gives a very good approximation of 
${\rm dim }S^2 =2$. This kind of analysis will be possible for more
general configurations.}.

Although we only considered the problem of 
the classical limit in this paper, 
finding a general construction method of the matrix regularization
is also a very important problem. 
In our formulation, coherent states introduce the notion of 
locality for matrix configurations.
This idea may help to construct the matrix regularization for 
a general manifold.
For instance, for a given manifold, 
one can first discretize the manifold to a fine lattice, 
and let the coordinates of each lattice cite 
to be diagonal elements of the matrices.
In order to determine the off-diagonal elements, 
one may refer to the representation
of the coordinate operators on 
the noncommutative plane in the basis 
of the coherent states.
If any matrix geometry looks locally like 
the noncommutative plane, 
the off-diagonal elements should be locally the same as those on 
the noncommutative plane. 
If so, by gluing those components
along the tangent directions, it will be possible to construct 
the $D$ Hermitian matrices which provide the 
matrix regularization.

We pursue these directions and 
hope to report these issues in the near future.

\section*{Acknowledgements}
This work was supported, in part, by Program to Disseminate Tenure Tracking System, MEXT, Japan.

\appendix 

\section{Canonical coherent states}
\label{Canonical coherent states}
In this Appendix, we consider the canonical coherent 
states and summarize their properties\cite{Gazeau:2009zz}. 
We start with the algebra of the creation-annihilation operators, 
\begin{align}
[\hat{a}, \hat{a}^{\dagger }]= 1.
\end{align}
The vacuum state is defined by 
\begin{align}
\hat{a}|0 \rangle =0, 
\end{align}
as well as $\langle 0 | 0 \rangle =1$.
We consider the Hilbert space given by
\begin{align}
{\cal H} = {\rm span}\{(\hat{a}^{\dagger})^n|0 
\rangle | n \in {\bf Z}, n\geq 0 \}.
\end{align}
The canonical coherent states are elements of
${\cal H}$ parametrized by 
a complex number $z \in {\bf C}$, and defined as
\begin{align}
|z \rangle = e^{z\hat{a}^{\dagger}-\bar{z}\hat{a}}|0 \rangle.
\label{def of can coh}
\end{align}
Let us introduce the Hamiltonian of the harmonic oscillator, 
\begin{align}
H = \hbar \left( \hat{a}^{\dagger} \hat{a} +\frac{1}{2} \right) , 
\end{align}
and its eigenstates
\begin{align}
H|n \rangle = E_n |n \rangle, \;\; \;\;
E_n = \hbar\left(n+\frac{1}{2} \right), \;\; \;\;
|n \rangle = \frac{1}{\sqrt{n!}}(\hat{a}^{\dagger})^n|0 \rangle 
\;\;\;\; (n=1,2, \cdots ).
\end{align}
The coherent states can be written in terms of $|n\rangle $ as
\begin{align}
|z \rangle = 
e^{-\frac{1}{2}|z|^2} \sum_{n=0}^{\infty} \frac{z^n}{\sqrt{n! }}
|n \rangle. 
\label{z and n}
\end{align}

The canonical coherent states have the following properties:
\begin{align}
&{\rm (i)} \;\; 
\langle z_1 | z_2 \rangle = e^{-\frac{|z_1|^2}{2}-\frac{|z_2|^2}{2}+
\bar{z}_1z_2},
\\
&{\rm (i\hspace{-.08em}i)} \;\;
\int \frac{d^2 z}{\pi }|z \rangle \langle z | = 1 
\;\;\; {\rm ( resolution \; of \; identity )},
\label{resolution of identity}
\\
&{\rm (i\hspace{-.08em}i\hspace{-.08em}i)}\;\;
\hat{a}|z \rangle   = z|z \rangle,
\\
&{\rm (i\hspace{-.08em}v)} \;\;
|z \rangle \; {\rm is \; the \; ground \; state \; of \; the \; 
shifted \; Hamiltonian, \; }
\nonumber\\
& \hspace{2cm} H(z) = \hbar \left( 
(\hat{a}^{\dagger}-\bar{z})(\hat{a}-z)+\frac{1}{2}
\right),
\\
&{\rm (v)} \;\;
|z\rangle {\rm \; saturates \; the \; uncertainty \; inequality }.
\end{align}
The first and the second properties can be derived 
from (\ref{z and n}) and 
the completeness relation $\sum_n |n \rangle \langle n| =1$.
The integration measure in (\ref{resolution of identity})
is defined as the flat measure for 
the real and imaginary parts of $z$, 
$\int d^2z := \int_{-\infty}^{\infty}d({\rm Re}z)  
\int_{-\infty}^{\infty}d({\rm Im}z)$.
The third, fourth and fifth properties 
follow from the fact that 
the unitary operator 
$U(z)= e^{z\hat{a}^{\dagger}-\bar{z}\hat{a}}$, which appears in 
Eq. (\ref{def of can coh}), is the translation operator 
$U^{\dagger}(z)\hat{a} U(z) = \hat{a}+z $.

\section{Bloch coherent states}
\label{Bloch coherent states}
Coherent states for the fuzzy sphere are called the
Bloch (spin) coherent states\cite{Gazeau:2009zz}\footnote{
See also \cite{Grosse:1994ed,Hammou:2001cc} for 
the description of fuzzy sphere using coherent states.}.
The fuzzy sphere is described by the $SU(2)$ Lie algebra:
\begin{align}
[L_i, L_j]= i \epsilon_{ijk}L_k.
\end{align}
We consider the spin $J$ representation, so that the $L_i$ are 
$(2J+1) \times (2J+1)$ matrices.
We introduce the standard basis for this representation space,
which satisfy
\begin{align}
&L_3|Jm\rangle = m| Jm\rangle,
\nonumber\\
&L_{\pm} |Jm\rangle = \sqrt{(J\mp m)(J\pm m+1)}|J m\pm 1 \rangle ,
\label{action of L}
\end{align}
where $L_{\pm}=L_1 \pm iL_2$, $m = -J, -J+1, \cdots, J$ and 
the states are normalized as $\langle Jm | Jm' \rangle = \delta_{mm'}$.
The Bloch coherent states are parametrized by the 
coordinates on $S^2$, $\Omega = (\theta, \phi)$, where 
$0\leq \theta \leq \pi$ and $0 <  \phi \leq 2\pi$, 
and defined by 
\begin{align}
| \Omega \rangle = 
e^{\frac{1}{2}\theta e^{i\phi }L_- 
-\frac{1}{2}\theta e^{-i\phi}L_+ }|JJ \rangle .
\label{def of bloch}
\end{align}
By using the Baker-Canbell-Hausdorff formula, they can also be 
written as 
\begin{align}
| \Omega \rangle =e^{z L_- }e^{-L_3 \log(1+|z|^2 )}e^{-\bar{z}L_+}
|JJ \rangle ,
\end{align}
where $z = \tan \frac{\theta }{2} e^{i\phi }$.
By acting the operators onto $|JJ \rangle $,
one can rewrite them as 
\begin{align}
 | \Omega \rangle &= \frac{1}{(1+|z|^2)^J}\sum_{m=-J}^J
z^{J-m}
\left(\!\!
\begin{array}{c}
2J \\
J+m
\end{array}
\!\!
\right)^{\frac{1}{2}} |J m \rangle
\nonumber\\
&=\sum_{m=-J}^J 
\left( \!\!
\begin{array}{c}
2J \\
J+m
\end{array}
\!\!
\right)^{\frac{1}{2}}
\left( 
\cos \frac{\theta }{2}
\right)^{J+m}
\left( 
\sin \frac{\theta }{2}
\right)^{J-m}
e^{i(J-m)\phi }| Jm \rangle.
\label{exp for Bloch}
\end{align}

The Bloch coherent states have the following properties:
\begin{align}
&{\rm (I)} \;\; 
\langle \Omega_1 | \Omega_2 \rangle 
= \left(
\cos \frac{\theta_1 }{2}
\cos \frac{\theta_2 }{2}
+e^{i(\phi_2 -\phi_1)}
\sin \frac{\theta_1 }{2}
\sin \frac{\theta_2 }{2}
\right)^{2J},
\\
&{\rm (I\hspace{-.08em}I)} \;\;
\frac{2J+1}{4\pi }
\int_{S^2} d\Omega 
|\Omega \rangle \langle \Omega | = 1
\;\;\; {\rm ( resolution \; of \; identity )},
\label{resolution of identity}
\\
&{\rm (I\hspace{-.08em}I\hspace{-.08em}I)}\;\;
y^{\mu}L_{\mu}|\Omega \rangle = J|y||\Omega \rangle ,
{\rm \; where \;} y^{\mu} {\rm \; are \; given \; by}
\\
& \hspace{1cm} y^1 = |y|\sin \theta \cos \phi,
 \hspace{0.5cm} y^2 = |y|\sin \theta \sin \phi,
 \hspace{0.5cm} y^3 = |y|\cos \theta,
\\
&{\rm (I\hspace{-.08em}V)} \;\;
|\Omega \rangle \; {\rm is \; the \; ground \; state \; of \; the 
\; Hamiltonian, \; }
\nonumber\\
& \hspace{2cm} H(y) :=\frac{1}{2} 
\left( \frac{1}{\sqrt{J(J+1)}}L_{\mu}-y_{\mu }
\right)^2,
\\
&{\rm (V)} \;\;
|\Omega \rangle {\rm \; minimizes \;  }
\sum_{\mu =1}^3 (\Delta L_{\mu })^2, 
{\rm \; where \;} \Delta L_{\mu } {\rm \; is \; the \; 
standard \; deviation \; of \;}L_{\mu }.
\end{align}
The first and second properties directly follow from 
(\ref{exp for Bloch}).
In property (I\hspace{-.08em}I), 
the integration measure is defined as 
$\int d\Omega = \int_0^{\pi} d\theta \int_0^{2\pi}d\phi 
\sin \theta $.
The remaining properties follow from the fact that 
the unitary operator $R(\Omega ):= 
e^{\frac{1}{2}\theta e^{i\phi }L_- 
-\frac{1}{2}\theta e^{-i\phi}L_+ } $ in 
Eq. (\ref{def of bloch}) satisfies 
$ R(\Omega )L_3 R^{-1}(\Omega )= \frac{y^{\mu}L_{\mu}}{|y|} $.

\section{Derivation of equation (\ref{distance squared}) }
\label{Derivation 1}
In this Appendix, we derive (\ref{distance squared}).
As a preliminary step, we first consider the following object:
\begin{align}
\langle 0, y_1 | H(y) | 0, y_2 \rangle
= \left( E_0(y_2) +\frac{1}{2}(y^{\mu}_2 -y^{\mu})^2 \right) 
\langle 0,y_1 |0,y_2 \rangle
+(y^{\mu}_2-y^{\mu})\langle 0,y_1 |
X_{\mu}-y_{2\mu}|0,y_2 \rangle.
\label{object1}
\end{align}
From (\ref{cauchy}), we obtain
\begin{align}
\lim_{N\rightarrow \infty} 
(y^{\mu}_2-y^{\mu}_1) \langle 0,y_1 |0,y_2 \rangle
= 0  \;\;\; (y_1, y_2 \in {\cal M}).
\end{align}
This implies that the coherent states at different points are orthogonal 
to each other:
\begin{align}
\lim_{N\rightarrow \infty} 
\langle 0,y_1 |0,y_2 \rangle
= \delta^{(D)} (y_1-y_2) \;\;\; (y_1, y_2 \in {\cal M}).
\label{orthogonality of coh st}
\end{align}
By applying (\ref{cauchy}) and (\ref{orthogonality of coh st}) to 
the object (\ref{object1}), we obtain
\begin{align}
\lim_{N\rightarrow \infty} 
\langle 0, y_1 | H(y) | 0, y_2 \rangle
= \frac{1}{2}(y^{\mu}_2 -y^{\mu})^2
\delta^{(D)} (y_1-y_2) \;\;\; (y_1, y_2 \in {\cal M}).
\label{lemma}
\end{align}

Now, let us compute 
\begin{align}
f(y + \epsilon_{\perp})= 
\lim_{N\rightarrow \infty} 
\min_{| \alpha \rangle \in {\cal H}^*}
\langle \alpha | H(y+ \epsilon_{\perp})
| \alpha \rangle,
\label{fperp}
\end{align}
where $y \in {\cal M}$ and $\epsilon_{\perp}$ is 
a normal vector. 
Let us denote by ${\cal H}_{\rm coh}$ the Hilbert space spanned by 
all the coherent states:
\begin{align}
{\cal H}_{\rm coh} = {\rm span} \{|0,y' \rangle | y' \in {\cal M} \}.
\end{align}
The total Hilbert space can be written as 
\begin{align}
{\cal H}= {\cal H}_{\rm coh} \oplus \tilde{\cal H}.
\end{align}
By definition, $\tilde{\cal H}$ is the Hilbert space which 
does not have any coherent state. 
We can show that when $|\epsilon_{\perp}|$ is sufficiently small, 
the minimum in (\ref{fperp}) is saturated by an element in 
${\cal H}_{\rm coh}$. 
Hence, in the computation of (\ref{fperp}), 
it suffices to consider the case where 
$|\alpha \rangle $ is an element of $ {\cal H}_{\rm coh}$.
Any element $| \alpha \rangle \in {\cal H}_{\rm coh}$ can be 
expanded by the coherent states as 
\begin{align}
| \alpha \rangle =
\int d^Dy' \alpha (y') |0, y' \rangle,
\label{coh st expansion}
\end{align}
where $\alpha(y')$ is a function which vanishes unless $y'\in {\cal M}$.
Since $| \alpha \rangle $ in (\ref{fperp}) is normalized as
$\langle \alpha | \alpha \rangle =1$, 
this gives a constraint on $\alpha(y')$.
By substituting the expansion (\ref{coh st expansion}) 
into (\ref{fperp}) and using (\ref{lemma}), we obtain
\begin{align}
f(y + \epsilon_{\perp})= \frac{1}{2}
\int d^Dy' |\tilde{\alpha} (y')|^2
(y'^{\mu}-y^{\mu}-\epsilon^{\mu}_{\perp})^2.
\label{last eq}
\end{align}
Here, $\tilde{\alpha}(y')$ is the large-$N$ limit of 
the function which saturates the minimum in (\ref{fperp}).
It satisfies $\int d^D y' |\tilde{\alpha}(y')|^2 =1$ and vanishes on the 
outside of ${\cal M}$.   
Obviously, the quantity on the right-hand side of 
(\ref{last eq}) is minimized when 
$\tilde{\alpha}(y')$ localizes at $y$. Thus, we finally 
obtain (\ref{distance squared}).

\section{Jacobi identity }
\label{Jacobi identity}
In this Appendix, we show that $W^{\mu \nu}(y)$ defined in
(\ref{poisson}) satisfies
\begin{align}
W^{\mu \nu}(y) \partial_{\mu} W^{\rho \sigma }(y)
+
W^{\mu \rho}(y) \partial_{\mu} W^{\sigma \nu }(y)
+
W^{\mu \sigma}(y) \partial_{\mu} W^{\nu \rho }(y)=0.
\label{Jacobi}
\end{align}
This is equivalent to the Jacobi identity for the Poisson bracket 
defined by $\{A, B \}= W^{\mu \nu}(\partial_{\mu}A) (\partial_{\nu}B) $.
We consider an arbitrary polynomial $\Phi(X)$ and define 
a  corresponding function by 
\begin{align}
\phi (y) = \lim_{N \rightarrow \infty}
\langle 0,y | \Phi(X)| 0, y \rangle. 
\end{align}
We can show that 
\begin{align}
\lim_{N \rightarrow \infty }
c \langle 0,y | [X^{\mu}, \Phi(X)]| 0,y \rangle 
= W^{\mu \nu} (y) \partial_{\nu }\phi (y).
\label{general jacobi}
\end{align}
If we put $\Phi = c[X^{\mu}, X^{\nu}]$ in the above equations, 
(\ref{Jacobi}) immediately 
follows from the Jacobi identity of the matrix commutators.

The relation (\ref{general jacobi}) can be shown as follows. 
The right-hand side of (\ref{general jacobi}) is the large-$N$ limit of 
\begin{align}
c \langle 0,y | [X^{\mu}, X^{\nu }] | 0, y \rangle 
\partial_{\nu } \langle 0,y | \Phi(X) | 0,y \rangle. 
\label{eq1 for Jacobi}
\end{align}
The derivative of $|0, y \rangle $ can be read off from the 
formula in the perturbation theory as
\begin{align}
\partial_{\nu }|0, y \rangle = \sum_{n\neq 0} 
\frac{|n, y \rangle \langle n, y | X_{\nu}|0,y \rangle  }
{E_n (y)- E_0 (y)}.
\end{align}
By substituting this, and using (\ref{useful}), we can see that 
the large-$N$ limit of (\ref{eq1 for Jacobi}) 
is equal to the left-hand side of (\ref{general jacobi}).

\section{Classical geometry for fuzzy torus}
\label{Classical geometry for fuzzy torus}
In this Appendix, we show that the function $f(y)$ associated with
the Hamiltonian (\ref{H for FT}) is given by (\ref{F for FT}).
First, note the following inequalities, 
\begin{align}
E_0(y) &= \min_{|\psi \rangle  \in {\cal H}^*} 
\langle \psi | H(y)| \psi \rangle
\nonumber\\
&\geq \frac{1}{2}\min_{|\psi \rangle  \in {\cal H}^*} 
 \langle \psi | (U-z)(U^{\dagger } -\bar{z}) | \psi \rangle
+ \frac{1}{2}\min_{|\psi \rangle  \in {\cal H}^*} 
 \langle \psi | (V-w)(V^{\dagger } -\bar{w}) | \psi \rangle 
\nonumber\\
& \geq 
\frac{1}{2}(1-|z|)^2 + \frac{1}{2}(1-|w|)^2.
\label{ineq 1}
\end{align}
The last inequality follows from the fact that the spectrum of any 
unitary matrix is contained in the unit circle on the complex plane.
On the other hand, for any state vector 
$|\alpha \rangle \in {\cal H}^*$, we have 
\begin{align}
E_0(y) \leq \langle \alpha | H(y) | \alpha \rangle. 
\label{ineq 2}
\end{align}
Hence, if there exists a state vector $|\alpha \rangle \in {\cal H}^* $ 
which satisfies
\begin{align}
\lim_{N\rightarrow \infty} \langle \alpha | H(y) | \alpha \rangle
= \frac{1}{2}(1-|z|)^2 + \frac{1}{2}(1-|w|)^2,
\label{condition for beta}
\end{align}
we can prove (\ref{F for FT}) from the two inequalities
(\ref{ineq 1}) and (\ref{ineq 2}).
In the following, we explicitly construct a state
$| \alpha \rangle $ which satisfies 
(\ref{condition for beta}).

We introduce the basis used in Eq. (\ref{clock shift}), 
\begin{align}
U|m \rangle = e^{i \theta }| m \rangle, \;\;\; 
V|m \rangle = | m+1 \rangle.
\end{align}
For convenience, we extend the range of $m$ to a set of all
integers by assuming the periodicity condition $|m+N \rangle =|m \rangle $.
With this notation, we introduce a state vector 
\begin{align}
| \alpha \rangle := 
\tilde{c} \sum_{m=-\left[\frac{N-1}{2} \right]}^{\left[\frac{N}{2} \right]} 
e^{- \frac{a}{2}\left(\frac{m}{N}-\frac{{\rm arg}z}{2\pi} \right)^2}
e^{-i ({\rm arg}w) m } |m \rangle,
\end{align}
where $a $ is a constant of ${\cal O}(\sqrt{N})$,
$\tilde{c}$ is a normalization
constant determined by $\langle \alpha | \alpha \rangle =1 $ and 
$[x]$ stands for the floor function defined by 
$[x]=\max \{m \in {\bf Z} | m \leq x  \}$.
In the large-$N$ limit, we can evaluate $\tilde{c}$ as
\begin{align}
\tilde{c}^{-2} = \sum_{m=-\left[\frac{N-1}{2} \right]}^{\left[\frac{N}{2} \right]} 
e^{- a \left(\frac{m}{N}-\frac{{\rm arg}z}{2\pi} \right)^2}
 \sim
N\int^{\infty }_{-\infty }d x
e^{- a \left(x-\frac{{\rm arg}z}{2\pi} \right)^2}
= N \sqrt{\frac{\pi }{a}},
\end{align}
where we have approximated the discrete sum by the 
integral\footnote{The integration range is naively given by 
$[-1/2, 1/2]$. However, since we assumed that $a={\cal O}(\sqrt{N})$,
the main contribution comes only from the neighborhood of the origin. 
Hence we can extend the range to $[-\infty , \infty]$}. 
By repeating similar calculations, we obtain
\begin{align}
\lim_{N\rightarrow \infty} 
\langle \alpha | U | \alpha \rangle = e^{i {\rm arg} z}, 
\;\;\;
\lim_{N\rightarrow \infty} 
\langle \alpha | V | \alpha \rangle = e^{i {\rm arg} w}.
\end{align}
This implies that $| \alpha \rangle $ satisfies 
(\ref{condition for beta}).

\section{Derivation of equation (\ref{f for moyal})}
\label{Derivation of equation}
In this Appendix, we derive (\ref{f for moyal}).
Up to the second order of the perturbation, 
the ground state energy is given by 
\begin{align}
E_{0,0} (y) = 
& \frac{\theta + \theta' }{2} +\frac{1}{2}\sum_{I=5}^D (y_I)^2
+ \langle 0, 0, y | H_1(y)+H_2(y)|0,0, y \rangle 
\nonumber\\
& + \sum_{(m,n )\neq (0, 0)}
\frac{\langle 0, 0, y | H_1(y)| m,n,y \rangle \langle m,n,y |
H_1(y)|0,0,y \rangle }{E_{0,0}^{(0)}(y)-E_{m,n}^{(0)}(y)}.
\label{perturbation}
\end{align}

We first construct the wave functions for
 the zeroth-order Hamiltonian $H_0 (y)$ 
in (\ref{Ham for infinite}).
Since $H_0 (y)$ is just the sum of two independent (shifted) 
harmonic oscillators, 
the wave function is given by the product of wave functions for each 
oscillator, 
\begin{align}
\Phi_{m,n} (x, \tilde{x}; y) 
= \Psi^{(\theta )}_m(x ;y^1, y^2)\Psi^{(\theta ')}_n(\tilde{x} ;y^3, y^4).
\label{wf}
\end{align}
$\Psi^{(\theta)}_m(x ;y^1, y^2)$ 
is the wave function of the $m$th excited state of 
the shifted harmonic oscillator, given by 
\begin{align}
\Psi^{(\theta )}_m(x ;y^1, y^2) = c_n h_n
\left(\frac{x-y_1 }{\sqrt{\theta} }\right)
e^{-\frac{1}{2\theta }(x-y^1)^2 +\frac{iy^2 x}{\theta } },
\end{align}
where $c_n =1/\sqrt{2^n n! \sqrt{\theta \pi }}$ and 
$h_n(\xi )$ is the Hermite polynomial defined by 
$h_n(\xi )= (-1)^n e^{\xi^2 }\left(\frac{d}{d\xi }\right)^n e^{-\xi ^2}$.
The operators $\hat{q}^A$ act on $\Phi_{m,n} (x, \tilde{x}; y)$
as follows:
\begin{align}
&(\hat{q}^1 \Phi_{m,n}) (x, \tilde{x}; y) = x\Phi_{m,n} (x, \tilde{x}; y),
\nonumber\\
&(\hat{q}^2 \Phi_{m,n}) (x, \tilde{x}; y) = -i\theta 
\frac{\partial }{\partial x}\Phi_{m,n} (x, \tilde{x}; y),
\nonumber\\
&(\hat{q}^3 \Phi_{m,n}) (x, \tilde{x}; y) = \tilde{x}
\Phi_{m,n} (x, \tilde{x}; y),
\nonumber\\
&(\hat{q}^4 \Phi_{m,n}) (x, \tilde{x}; y) = -i\theta' 
\frac{\partial }{\partial \tilde{x}}\Phi_{m,n} (x, \tilde{x}; y).
\end{align}
By using the  wave functions (\ref{wf}), 
we can obtain
\begin{align}
&\langle 0,0,y | e^{ik \cdot \hat{q}} | m,n,y \rangle 
= 
\frac{c_mc_n}{c_0^2}
(i\sqrt{\theta} (k_1-ik_2))^m 
(i\sqrt{\theta'} (k_3-ik_4))^n
e^{-\frac{\theta}{4}k_a^2 -\frac{\theta'}{4}k_{\alpha }^2
+i k\cdot y},
\nonumber\\
&\langle 0,0,y | \{ \hat{q}_a -y_a,  e^{ik \cdot \hat{q}} \} | m,n,y \rangle 
= i \left[
k_a \theta - \frac{2m  \chi_a}{ k_1-ik_2 } 
\right]
\langle 0,0,y | e^{ik \cdot \hat{q}} | m,n,y \rangle,
\nonumber\\
&\langle 0,0,y | \{ \hat{q}_{\alpha} -y_{\alpha },  
e^{ik \cdot \hat{q}} \} | m,n,y \rangle 
= i \left[
k_{\alpha } \theta' - \frac{2n  \chi_{\alpha }}{ k_3-ik_4 } 
\right]
\langle 0,0,y | e^{ik \cdot \hat{q}} | m,n,y \rangle,
\label{q-a times e}
\end{align}
where $\chi_a$ and $\chi_{\alpha }$ are constants 
given by $\chi_1=\chi_3 =1$, $\chi_2=\chi_4=-i$.
By using (\ref{q-a times e}), we can compute the correction terms in 
(\ref{perturbation}).
In the classical limit where $\theta, \theta' \rightarrow 0$, 
the leading behaviors are given as follows:
\begin{align}
&\langle 0,0,y | H_1(y)| 0,0,y \rangle 
\sim 
- y^I \phi_I (y),
\nonumber\\
& \frac{1}{\sqrt{\theta }}
\langle 0,0,y| H_1(y)|1,0,y \rangle 
\sim \frac{1}{\sqrt{2}}
\left\{
\phi_1(y)-i\phi_2(y)-y^I(\partial_1\phi_I(y)-i\partial_2\phi_I(y))
\right\},
\nonumber\\
&\frac{1}{\sqrt{\theta' }}
\langle 0,0,y| H_1(y)|0,1,y \rangle 
\sim \frac{1}{\sqrt{2}}
\left\{
\phi_3(y)-i\phi_4(y)-y^I(\partial_3\phi_I(y)-i\partial_4\phi_I(y))
\right\},
\nonumber\\
&\langle 0,0,y | H_2(y)| 0,0,y \rangle 
\sim
\frac{1}{2}(\phi^{\mu}(y))^2.
\label{perturbative terms}
\end{align}
In deriving these, 
the following approximation formula for the 
delta function is useful:
\begin{align}
\int \frac{d^4 k}{(2\pi )^4}
e^{-\frac{\theta}{4}k_a^2 -\frac{\theta'}{4}k_{\alpha }^2
+i k\cdot (y-x)}
&= \frac{1}{\pi^2 \theta \theta' }
e^{-\frac{1}{\theta}(y_a-x_a)^2-\frac{1}{\theta' }(y_{\alpha}-x_{\alpha})^2}
\sim \delta^{(4)}(x-y).
\end{align}
By substituting (\ref{perturbative terms}) into 
(\ref{perturbation}), we finally obtain 
(\ref{f for moyal}).

\end{document}